\documentclass[twocolumn,showpacs,10pt]{revtex4-1} 
\usepackage{amsmath,amssymb,graphicx,comment}

\newcommand{\uu}{\mathcal{\hat U}}

\newcommand{\hc}{\text{H.c.}}

\newcommand{\vac}{\text{vac}}

\newcommand{\ket}[1]{| #1 \rangle}

\begin{document}


\title{High efficiency in mode selective frequency conversion}

\author{Nicol\'as Quesada and J.E. Sipe}
\affiliation{McLennan Physical Laboratories, University of Toronto, Toronto, ON, M5S 1A7, Canada}

\begin{abstract}
Frequency conversion (FC) is an enabling process in many quantum information protocols. 
Recently, it has been observed that upconversion efficiencies in single-photon, mode-selective FC are limited to around 80\%.
In this letter we argue that these limits can be understood as time-ordering corrections (TOCs) that modify the joint conversion amplitude of the process. Furthermore we show, using a simple scaling argument, that recently proposed cascaded FC protocols that overcome the aforementioned limitations act as ``attenuators'' of the TOCs. 
This observation allows us to argue that very similar cascaded architectures can be used to attenuate TOCs in photon generation via spontaneous parametric down-conversion.
Finally, by using the Magnus expansion, we argue that the TOCs, which are usually considered detrimental for FC efficiency, can also be used to increase the efficiency of conversion in partially mode selective FC.  
\end{abstract}

\maketitle

Nonlinear photonic materials provide some of the most advanced platforms for manipulating the frequency and spectral profile of photons. 
This manipulation is typically achieved using a process known as frequency conversion (FC)\cite{kumar90,huang92}. 
In the version of FC we consider, two photons (one of them typically coming from a bright classical field) are fused into another photon with a higher energy by the process of sum frequency generation. 
FC has several important applications, including photon detection\cite{lang05}, and the establishment of compatibility between sources and quantum memories \cite{donohue15}. 
FC can also be used to modify in a controlled manner the properties of weak signals or single photons, a useful component of several quantum information processing protocols that harness the infinite dimensional Hilbert space structure of the frequency degree of freedom of photons \cite{brecht15,eckstein11,raymer12,reddy14}. A set of orthogonal frequency amplitude functions for the photon(s) (henceforth referred as ``modes'') provides a natural basis in which to encode information in this Hilbert space \cite{brecht15}, and FC provides a natural way to do controlled operations in this Hilbert space \cite{eckstein11}.
For all applications of FC it is important to have conversion efficiency near unity, and for controlled operations it is also important to have mode selectivity.  
In the limit of very short crystals, or equivalently very long pulses (effectively CW fields) it has been shown that 100\% FC is achievable \cite{clark13,donohue15}. 
Nevertheless in these limits most mode selectivity is lost.
In this letter, we examine limitations to highly efficient mode selective FC. 
We show below that these limitations are due to time-ordering corrections (TOCs) that appear because the interaction picture Hamiltonian that describes the $\chi_2$ (or $\chi_3$) interaction between the different fields does not commute with itself at different times. 
Our study allows us to separate very cleanly the ``ideal'' operation of an FC device from the ``undesirable'' effects of time ordering. 
Thus, we can explicitly write the operation of an FC gate, in a language very close to the one used in quantum information, as a unitary operation $ \uu_{\text{FC}}=\exp(\hat \Omega_1+\hat \Omega_2+\hat \Omega_3+ \ldots)$ that consists of a desired generator $\hat \Omega_1$ and TOCs $\hat \Omega_2, \hat \Omega_3, \ldots $ that modify the operation of the gate. 
This separation is obtained using the Magnus expansion (ME)\cite{nico14}.  
Using the ME allows us to understand the scaling of the TOCs as a function of the energy of the classical pump pulse. 
Based on this model, we propose a new scheme to achieve highly efficient (and partially mode selective) FC, harnessing the TOCs that until now have been undesirable; this model also allows us to explain, using simple scaling arguments, why the double pass scheme introduced by Reddy \emph{et al.} \cite{reddy14} succeeds in achieving high efficiency FC, and to extend these ideas to the case of photon generation using spontaneous parametric down-conversion or four wave mixing.

We begin with the Hamiltonian describing a second-order nonlinear optical FC process in the undepleted pump approximation, occurring in a quasi one-dimensional structure of length $L$ and characterized by a nonlinear coefficient $\chi_2$ \cite{nico14},
\begin{align}\label{HI}
\hat{H}_{I}(t)=&-\hbar \varepsilon \int d\omega _{p}d\omega _{a}d\omega _{b} \big(e^{i\bar{ \Delta}t}{\Phi}( \overline{ \Delta k}(\omega _{a},\omega _{b},\omega _{p})L/2) \nonumber \\ 
& \quad \quad {\alpha}(\omega _{p})\hat{a}(\omega _{a})\hat{b}^{\dagger }(\omega_{b})+\hc\big),
\end{align}
where $\omega_p$ refers to a pump frequency, and the shape of the classical pump function $\alpha(\omega_p)$ is taken to be a Gaussian,
\begin{align}
\alpha(\omega_p)=\tau e^{-\tau ^{2}\delta \omega_p^{2}}/\sqrt{\pi }, \quad \delta \omega_p=\omega _{p}-{\bar \omega}_{p},
\end{align}
where $\bar \omega_p$ identifies the center pump frequency; $\tau$ identifies the duration of the pump pulse, and we specify its energy by $U_0$. In the FC process frequency components $\omega_a$ can be destroyed and frequency components $\omega_b$ can be created; associated with this are bosonic destruction and creation operators $\hat a(\omega_a)$ and $\hat b^\dagger(\omega_b)$ respectively. We assume that the phase-matching function (PMF) $\Phi(x)$, where its argument $\overline{\Delta k}(\omega_a,\omega_b,\omega_p)\equiv k_b(\omega_b)-k_a(\omega_a)-k_p(\omega_p)\pm 2\pi/\Lambda$ (including the quantity $2\pi/\Lambda$ only if periodic poling for quasi phase matching is performed on the device), and the $k_i(\omega_i)$ indicate the dispersion relations of the modes involved, restricts the destroyed and created photons to be in nonoverlapping frequency regions, and/or of different mode profiles or polarizations; then we can take $[\hat a(\omega_a),\hat b^\dagger(\omega_b)]=0$. We reference the frequencies $\omega_a$ and $\omega_b$ to center frequencies $\bar \omega_a$ and $\bar \omega_b$ respectively, defined so that energy and momentum conservation are exactly satisfied for the center frequencies, 
\begin{align}\label{zero}
\bar \omega _{b}-\bar \omega _{a}-\bar \omega _{p}=0 \text{  and  } \overline{\Delta k}(\bar \omega_a,\bar \omega_b,\bar \omega_p)=0. 
\end{align}
Finally, $\bar{\Delta}=\omega _{b}-\omega _{a}-\omega _{p}$, and
\begin{align}
\varepsilon =2L\chi _{2}\sqrt{\frac{\sqrt{2}U_{0} \pi {\bar \omega} _{b}{\bar \omega} _{a}}{\sqrt{\pi}(4\pi)^{3}\epsilon _{0}Ac^{3}n_{a}({\bar \omega} _{a})n_{b}({\bar \omega} _{b})n_{c}({\bar \omega} _{c}) \tau }}
\end{align}
is a dimensionless constant that characterizes the strength of the interaction;  the $n_i (\bar \omega_i)$ are the indices of refraction  at the central frequencies $\bar \omega_i$, and $A$ is the effective overlap area of the spatial fields.

\begin{figure}
\centering
\includegraphics[width=0.40 \textwidth]{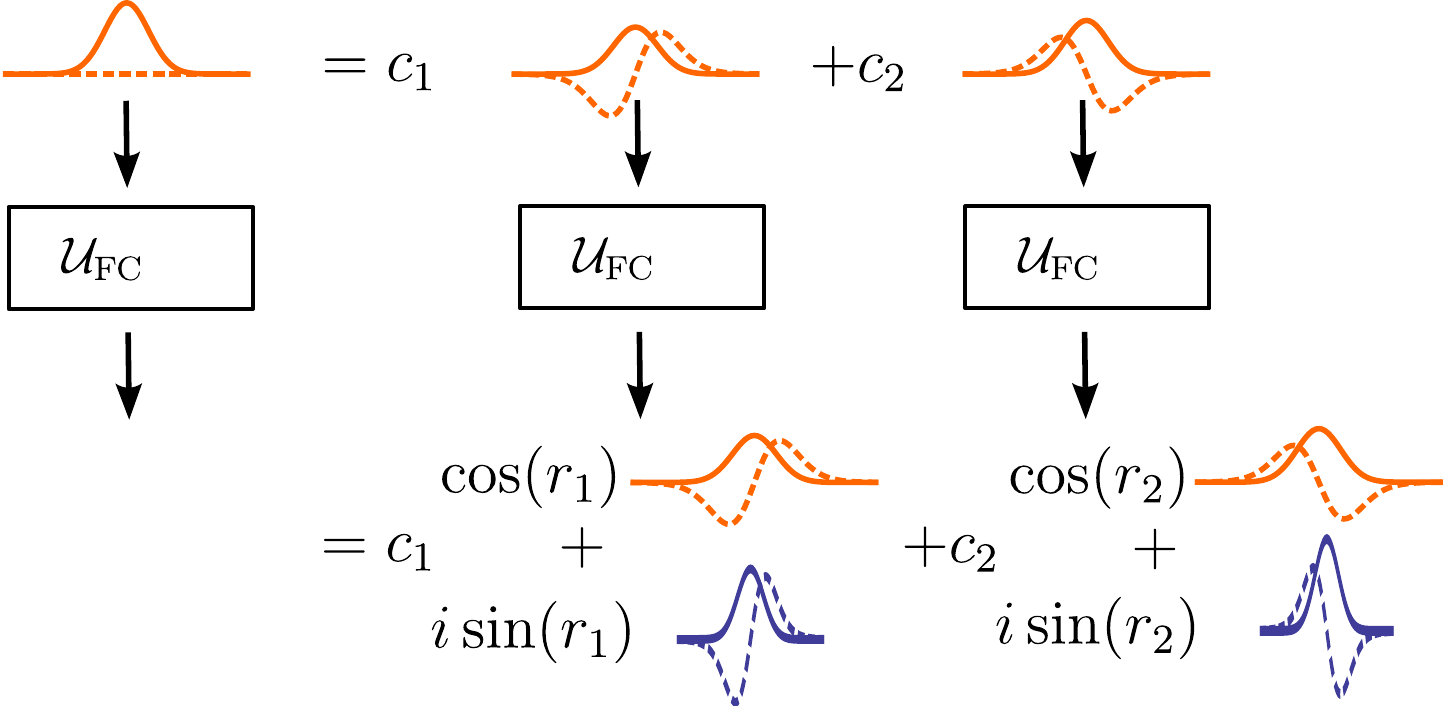}
\caption{\label{scheme} Schematic representation of a frequency converter. The device has a complete set of input Schmidt modes (represented in orange); an arbitrary input single photon can thus be represented in this basis with expansion coefficients $c_\theta$. As explained earlier \cite{eckstein11,ben11}, the frequency converter acts as a beam splitter of reflectivity $\sin(r_\theta)$ in each of the Schmidt modes connecting input modes to output modes represented in blue. Note that the mode profiles can in general have complex amplitudes and thus we use continuous/dashed lines for their real/imaginary parts.}
\end{figure}

Ignoring TOCs (as it is implicitly done in \cite{eckstein11}), the unitary evolution operator connecting states at $t \to -\infty$ to states at $t\to \infty$ is  
\begin{subequations}
\begin{align}\label{uwrong}
\uu_1&=e^{\hat \Omega_1}=e^{-\frac{i}{\hbar} \int_{-\infty}^{\infty} dt \hat H_I(t)}\\
&=e^{-2\pi i \int d\omega_a d\omega_b  \left(\bar{J_1}(\omega_a,\omega_b) \hat a(\omega_a) \hat b^\dagger(\omega_b)+\hc\right)}, 
\end{align}
\end{subequations}
where the quantity $\bar J_1$ is, to this level of approximation, the joint conversion amplitude (JCA) and is given by
\begin{align*}
\bar{J}_1(\omega_a,\omega_b)=- \varepsilon \alpha(\omega_b-\omega_a) {\Phi}( \overline{ \Delta k}(\omega _{a},\omega _{b},\omega _{b}-\omega_a)L/2). 
\end{align*}
To understand how this operator transforms a given input state, let us first introduce the Schmidt decomposition of the function $\bar{J_1}$
\begin{align}\label{schmidt}
-  \bar{J}_1(\omega_a,\omega_b)=\sum_{\theta} \frac{r_\theta(\varepsilon)}{2 \pi} k_{\theta}^*(\omega_a) l_{\theta}(\omega_b), \ r_\theta(\varepsilon)=\varepsilon \tilde r_\theta,
\end{align}
where the Schmidt functions $k_\theta(\omega)$ ($l_\theta(\omega)$) are assumed to form a complete and orthonormal set, and the positive quantities $r_{\theta}(\varepsilon)=\varepsilon \tilde r_\theta$ are the Schmidt numbers of the function $\bar{J_1}$. 
In Eq. (\ref{schmidt}) we have explicitly used  the fact that $\varepsilon$ is only a \emph{multiplicative constant} in this approximation of the JCA; accordingly, the Schmidt numbers are \emph{linear functions} of this quantity and the Schmidt functions are \emph{independent} of it. 
Because of the orthonormality of the functions $k_\theta(\omega)$ and $l_\theta(\omega)$, we can write
\begin{align}\label{fact1}
\uu_1=e^{i\sum_{\theta}  r_{\theta}(\varepsilon) (\mathcal{\hat A}_{\theta} \mathcal{\hat B}_{\theta}^\dagger+\hc)  }=\bigotimes_{\theta} e^{i  r_{\theta}(\varepsilon)  (\mathcal{\hat A}_{\theta} \mathcal{\hat B}_{\theta}^\dagger+\hc)},\\
\mathcal{\hat A}_{\theta}=\int d\omega_a k^*_\theta(\omega_a) \hat a(\omega_a), \ \mathcal{\hat B}_{\theta}=\int d\omega_b l^*_\theta(\omega_b) \hat b(\omega_b),
\end{align}
where we introduced broadband Schmidt operators $\mathcal{\hat A}_{\theta}$ and $\mathcal{\hat B}_{\theta}$ for fields $a$ and $b$. These operators transform in the expected way under $\uu_1$; for example,
\begin{align}
\uu_1 \mathcal{\hat A}_{\theta} \uu_1^\dagger&=\cos(  r_{\theta}(\varepsilon) ) \mathcal{\hat A}_{\theta} - i \sin(r_{\theta}(\varepsilon) ) \mathcal{\hat B}_{\theta} \label{trans1a}.
\end{align}
Let us consider how the unitary in Eq. (\ref{uwrong}) transforms a single photon of the form, 
\begin{subequations}\label{transgen2}
\begin{align}
\ket{1_{g(\omega_a)}}&=\int d\omega_a g(\omega_a) \hat a^{\dagger}(\omega_a)\ket{\vac}\label{trans1}\\
&=\sum_{\theta} \underbrace{\left(\int d\omega_a  k_\theta^*(\omega_a) g(\omega_a) \right)}_{\equiv c_\theta} \mathcal{\hat A}_\theta^\dagger \ket{\vac} \label{trans2}.
\end{align}
\end{subequations}
The amplitude $g$ is assumed to be normalized according to $\int d\omega |g(\omega)|^2=1$, and in the last equation we used the completeness of the set $\{ k_\theta(\omega_a) \}$. 
We can now use Eq. (\ref{trans1a}) and Eq. (\ref{trans2}) to study the effect of the unitary on the single photon state (\ref{trans1}), as sketched in Fig. \ref{scheme},
\begin{align}\label{transgen3}
\uu_1 \ket{1_{g(\omega_a)}}=\sum_{\theta}  c_\theta  
\left(\cos(r_{\theta}(\varepsilon)) \mathcal{\hat A}_{\theta}^\dagger + i \sin(r_{\theta}(\varepsilon)) \mathcal{\hat B}_{\theta}^\dagger \right)\ket{\vac} 
\end{align}
In particular, if a single photon with frequency profile $k_{\theta'}(\omega_a)$ is initially prepared, it will be upconverted to a photon with profile $l_{\theta'}(\omega_b)$  with probability amplitude $i \sin(\varepsilon r_{\theta'})$. One very interesting case of the last equation is when the JCA is separable $\bar J_1(\omega_a,\omega_b)=\varepsilon r_{0} k_{0}(\omega_a) l^{*}_{0}(\omega_b)$ and thus has only one nonzero Schmidt number in Eq. (\ref{schmidt}). Then one can selectively upconvert only Schmidt function $0$ and leave the rest unchanged; this is precisely the ideal operation of a quantum pulse gate (QPG) as introduced by Eckstein \emph{et al.} \cite{eckstein11}.
However, the $\uu_1$ in Eq. (\ref{uwrong}) is not the correct unitary operator since very generally $[\hat H_I(t),\hat H_I(t')]\neq 0$, and thus Eq. (\ref{uwrong}) should be premultiplied by the time-ordering operator $\mathcal{T}$ \cite{shankar12}, as has been mentioned in this context \cite{eckstein11,ben11} but not worked out in the needed detail. One very special an important case where Eq. (\ref{uwrong}) does hold is when the PMF is flat $\Phi(\overline{\Delta k} (\omega_a,\omega_b,\omega_p)) \approx 1$ since, as shown by Donohue \emph{et al.} \cite{donohue15}, $[\hat H_I(t),\hat H_I(t')]=0$ in that case. 

Recently  we have shown that the ME provides a particularly appealing way of approximating the time evolution operator for time dependent quadratic Hamiltonians \cite{nico14} like the one in Eq. (\ref{HI}). Using the ME, the correct unitary time evolution operator is\cite{nico14}
\begin{align}\label{ucorrect}
\uu_{\text{FC}}=\mathcal{T}e^{-\frac{i}{\hbar} \int_{-\infty}^{\infty} dt \hat H_I(t)}=e^{\hat \Omega_1+\hat \Omega_2+ \hat \Omega_3+\ldots}.
\end{align}
In the last equation, the odd order terms are beam-splitter-like operators
\begin{align}\label{odd}
\hat \Omega_{2n+1}&=\frac{2 \pi}{ i} \int d\omega_a d\omega_b \big( \bar{J}_{2n+1}(\omega_a,\omega_b) \hat a(\omega_a) \hat b^\dagger(\omega_b)+\hc \big) 
\end{align}
and the even order terms are rotation-like operators
\begin{align}\label{even}
\hat \Omega_{2n}=\frac{2 \pi}{ i}\sum_{c=a,b} \int d\omega_c d\omega_c' \bar{G}^c_{2n}(\omega_c,\omega_c') \hat c^\dagger(\omega_c) \hat c(\omega_c').
\end{align}
The functions $\bar{J}_{2n+1}, \bar{G}_{2n}$ can be reduced to integrals of the PMF and pump functions \cite{nico14}.

One can break the RHS of Eq. (\ref{ucorrect}) into a unitary involving only beam-splitter terms and one involving only rotation terms $\uu_{\text{FC}}=\uu_{\text{RA}}\uu_{\text{RB}}\uu_{\text{BS}}$ where $\uu_{\text{RA}}$ and $\uu_{\text{RB}}$ are generated by operators such as Eq. (\ref{even}) and $\uu_{\text{BS}}$ is generated by operators such as Eq. (\ref{odd}). The lowest order example of this factorization is obtained by using the Baker-Campbell-Hausdorff formula
\begin{align}\label{fact}
\uu_{\text{FC}}= \left(\uu_{\text{RA}}\uu_{\text{RB}}\right)  \left(\uu_{\text{BS}}\right) \approx e^{\hat \Omega_2} \ e^{\hat \Omega_1+\hat \Omega_3+\frac{[\hat \Omega_1,\hat \Omega_2]}{2}},
\end{align}
where
\begin{align*}
\frac{[\hat \Omega_1,\hat \Omega_2]}{2}=& (2 \pi)\int d \omega_a d \omega_b (\bar{K}_3(\omega_a,\omega_b) \hat a(\omega_a) \hat b^\dagger(\omega_b)-\hc)\\ 
\frac{\bar{K}_3(\omega_a,\omega_b)}{\pi}=&  \int d \omega \big(\bar{J}_1(\omega_a,\omega) \bar{G}_2^b(\omega_b,\omega) - \bar{G}_2^a(\omega,\omega_a) \bar{J_1}(\omega,\omega_b)\big).
\end{align*}
The factorization Eq. (\ref{fact}) is useful because we can write the unitary for FC as an entangling operation for fields $\hat a$ and $\hat b$ postmultiplied by local unitaries. These local unitaries cannot alter the entanglement between $a$ and $b$, and because they involve rotation operators of the form given in Eq. (\ref{even}), they cannot change the total number of photons in fields $a$ and $b$. Thus the unitary $\uu_{\text{BS}}$ contains all the information necessary to calculate the probability of upconversion. The JCA associated with $\uu_{\text{BS}}$ is now
\begin{align}\label{newj}
\bar J=\underbrace{\bar J_1}_{\propto \varepsilon}+\underbrace{\bar J_3+i\bar K_3}_{\propto \varepsilon^3}+\ldots
\end{align}
We can introduce the Schmidt decomposition of $\bar J$ just as was done in Eq. (\ref{schmidt}), except that now, because of the TOCs, the Schmidt functions depend on $\varepsilon$ \cite{reddy15a} \emph{and} the Schmidt numbers are nonlinear functions of $\varepsilon$; thus we must replace
\begin{subequations}\label{subsrules}
\begin{align}
k_{\theta}(\omega_a) \to k_{\theta}(\omega_a;\varepsilon), &\quad l_{\theta}(\omega_b) \to l_{\theta}(\omega_b;\varepsilon),\\
\mathcal{\hat A}_{\theta}\to \mathcal{\hat A}_\theta(\varepsilon), & \quad \mathcal{\hat B}_{\theta}\to \mathcal{\hat B}_\theta(\varepsilon),
\end{align}
\end{subequations}
in Eq. (\ref{fact1}). Note that Eqs (\ref{fact1},\ref{trans1},\ref{transgen3}) will still hold after the rules (\ref{subsrules}) are applied.
In principle 100\% efficiency FC could be achieved were $r_{i}( \varepsilon)=\pi/2$ for some $i$ and $\varepsilon$ and the single photon $\mathcal{\hat A}^\dagger_{i}(\varepsilon)\ket{\vac}=\int d\omega_a k_\theta(\omega_a;\varepsilon)\hat a^\dagger(\omega_a)\ket{\vac}$ sent to the frequency converter. 
Of course, this could be hard to attain experimentally, since TOCs make the JCA a complicated function of $\omega_a$ and $\omega_b$, and certainly introduce an imaginary component (see Eq. (\ref{newj})); thus the $k_{\theta}(\omega_a;\varepsilon)$ can become complicated as well.

Another way of achieving highly efficient FC is by eliminating the TOCs. To this end note that the $n^\text{th}$ Magnus term $\hat \Omega_n$ scales with $\varepsilon^n$. Now let us imagine that instead of sending our single photon once through an FC device together with a pump pulse of energy $U_0$, the effect of which is characterized by $\uu_{\text{FC}}(\varepsilon)$, we send it sequentially through $N$ copies of the original FC device, each with the same pump pulse shape but with pump energy $U_0/N^2$ (equivalently scaling $\varepsilon$ by $1/N$). Then the time evolution operator is
$(\uu_{\text{FC}}(\varepsilon/N))^N=e^{\hat \Omega_1+\hat \Omega_2/N+ \hat \Omega_3/N^2+\ldots};$
thus by having more than one FC device the TOCs can be reduced. This observation neatly explains the results of Reddy \emph{et al.}\cite{reddy14}, where highly efficient-mode selective FC is achieved using cascaded non-linear crystals. One complication with this scheme is the fact that it requires multiple FC devices, which typically will imply stabilizing and controlling more complex interferometers. 
Note that the argument just presented makes no assumptions about the nature of the time dependent Hamiltonian that generated the Magnus terms; thus, one can also use cascaded nonlinear devices to attenuate the TOCs in photon generation using spontaneous parametric down-conversion (SPDC) as well. As noted in \cite{nico14} the TOCs associated with SPDC have a very strong resemblance to the ones present in FC, and in particular the Magnus terms in both processes have exactly the same scaling with respect to the energy of the pump pulse. 

Let us now consider the possibility of achieving near unit efficiency FC by harnessing the TOCs that until now have been undesirable. That is, can we turn a bug into a feature? A full analysis would of course have to be done including the $\varepsilon^5$ and possibly higher terms in Eq. (\ref{newj}), to either take their effects into account or determine that they are negligible for the parameter space being considered. This involves only straightforward algebra and integration, but we defer a complete study to a later communication. 
Here we simply restrict ourselves to Eq. (\ref{newj}) and show how complete conversion can be achieved within this approximation, and what is the physics associated with this possibility. We begin by considering for simplicity that the JCA is separable and Gaussian if TOCs are ignored. This can be achieved by engineering a Gaussian PMF \cite{dosseva14}. Gaussian functions can also be used as a simple approximation to the more common sinc function one would encounter for a uniform nonlinearity in over a region of length $L$. One can match the FWHM of $e^{-\gamma x^2}$ to the FWHM of sinc($x$) by setting $\gamma\approx 0.193$. Assuming a Gaussian PMF, $\Phi(x)=\exp(-\gamma x^2)$ , we expand the phase mismatch around the central frequencies of the fields involved to get
\begin{align}\label{phms}
\Phi(\overline{\Delta k}(\omega_a,\omega_b,\omega_p)L/2)&=e^{-(s_b \delta \omega_b-s_a \delta \omega_a-s_p \delta \omega_p)^2},
\end{align}
where $s_i=\sqrt{\gamma} L/(2 v_i)$, $\frac{1}{v_i}=\frac{d k_i(\omega_i)}{d\omega_i}|_{\omega_i=\bar \omega_i}$ is the inverse group velocity of the $i^{\text{th}}$ field, and we have assumed that group velocity dispersion is negligible.
Under these approximations, ignoring TOCs the JCA is
\begin{align}\label{j1}
&\bar J_{1}( \omega _{a}, \omega _{b}) = -\varepsilon \tau \ e^{2\mu^2 \delta \omega_a \delta \omega_b -\mu_a^2 \delta \omega_a^2 -\mu_b^2 \delta \omega_b^2}/\sqrt{\pi}, \\
&\mu ^{2}=\tau ^{2}+(s_{p}-s_{a})(s_{p}-s_{b}), \ \mu_{a,b}^{2}=\tau ^{2}+(s_{p}-s_{a,b})^{2}.  \nonumber
\end{align} 
The necessary and sufficient condition for a separable Gaussian $\bar J_1$ is $\mu=0$. This will give an ideal QPG, 
\begin{align}\label{1st}
\uu_1&=e^{i \int d\alpha d\beta \left(\tilde{r_0} \varepsilon f(\alpha) f(\beta) \hat a(\alpha) \hat b^\dagger(\beta) +\hc\right)}, \\
\tilde r_0&=\frac{\sqrt{2} \pi \tau}{\sqrt{ \mu_a \mu_b}}, \ \alpha=\mu_a \delta \omega_a, \ \beta=\mu_b \delta \omega_b,\ f(x)=\frac{e^{-x^2}}{\sqrt[4]{\pi/2}},\nonumber
\end{align}
for the photon $\ket{1_{f(\alpha)}}=\mathcal{\hat A}_0^\dagger\ket{\vac}=\int d\alpha f(\alpha) \hat a^\dagger(\alpha)\ket{\vac}$, mapping it to the photon $\ket{1_{f(\beta)}}=\mathcal{\hat B}_0^\dagger\ket{\vac}=\int d\beta f(\beta) \hat b^\dagger(\beta)\ket{\vac}$.
In Eq. (\ref{1st}) we have used the fact that the Schmidt functions of (\ref{j1}) with $\mu=0$ are 
$k_{0}(\omega_a)=\sqrt{\mu_a}f^*(\alpha)$  and  $l_0(\omega_b)=\sqrt{\mu_b} f(\beta)$.
To achieve $\mu=0$ and an ideal QPG one can envision two strategies. In the first a fixed pump pulse of duration $\tau$ is assumed and we seek a material where the product of the relative inverse group velocities of the fields satisfy the constraint $0=\mu^2=\tau ^{2}+(s_{p}-s_{a})(s_{p}-s_{b})$. A necessary condition for this to happen is that $s_a<s_p<s_b$ or $s_b<s_p<s_a$. If $\mu=0$ is so achieved, $\bar{J}_1$ is separable and one can in principle upconvert photons of temporal duration $\mu_a>\tau$ centered around $\bar{\omega_a}$ to photons with temporal duration $\mu_b>\tau$ centered around $\bar{\omega_b}$. A second strategy is to consider a given material with parameters  $s_a, s_b$ and $s_p$. Then $\mu=0$ is achieved by choosing a pump pulse of duration $\tau=\sqrt{-(s_{p}-s_{a})(s_{p}-s_{b})}>0$, which will convert photons of duration $\mu_a$ to photons of duration $\mu_b$. Note that in both strategies the duration of the photons that can be transduced is always larger than the one of the classical pump pulse (see Eq. (\ref{j1})). 

Finally, let us mention that mode selectivity is lost either for a flat PMF ($s_a,s_b,s_p \ll \tau$) or for a very long pulse ($\tau \gg s_a,s_b,s_p$), for $\mu_a,\mu_b,\mu \approx \tau$ and the Schmidt number $S=\mu_a \mu_b/\sqrt{\mu_a^2 \mu_b^2-\mu^4}$ is very large. Nonetheless in both cases it is possible to achieve high conversion efficiency \cite{lang05,donohue15,clark13}.

As soon as one starts to approach an FC efficiency near unity, one needs to include the TOCs.
The unitary connecting inputs and outputs is given by Eq. (\ref{fact}). The efficiency of the conversion is now solely governed by the $\uu_{\text{BS}}$ term in Eq. (\ref{fact}), and the associated JCA  in Eq. (\ref{newj}). The amplitudes in Eq. (\ref{newj}) can be related to the terms of the TOCs for SPDC \cite{nico15,queth15} via
$\bar J_1(\delta \omega_a,\delta \omega_b) =  J_1(-\delta \omega_a,\delta \omega_b)$, 
$\bar J_3(\delta \omega_a,\delta \omega_b) = - J_3(\delta \omega_a,-\delta \omega_b)$, 
$\bar K_3(\delta \omega_a,\delta \omega_b) =  K_3(-\delta \omega_a,\delta \omega_b)$,
where the unbarred quantities refer to the TOCs for SPDC. Note that the TOCs will typically generate more Schmidt functions; even if $\bar J_1$ is separable, $\bar J$ in Eq. (\ref{newj}) will not be for $\varepsilon \nleq 1$. 
For the very simple Gaussian PMF, and assuming $\mu=0$, it turns out that the Schmidt functions of the time ordered corrected JCA, $\bar J$ in Eq. (\ref{newj}), are only functions of $\tilde r_0 \varepsilon$. This quantity is the Schmidt number of the Schmidt function of interest in the $\varepsilon \ll 1$ limit (see Eq. (\ref{schmidt})). Also note that $\tilde r_0$ parametrizes the departure of the separable $\bar J_1$ from a circle. For a perfectly round JCA it takes the maximum value $\tilde r_0=\pi$.
\begin{figure}
\centering
\includegraphics[width=0.5\textwidth]{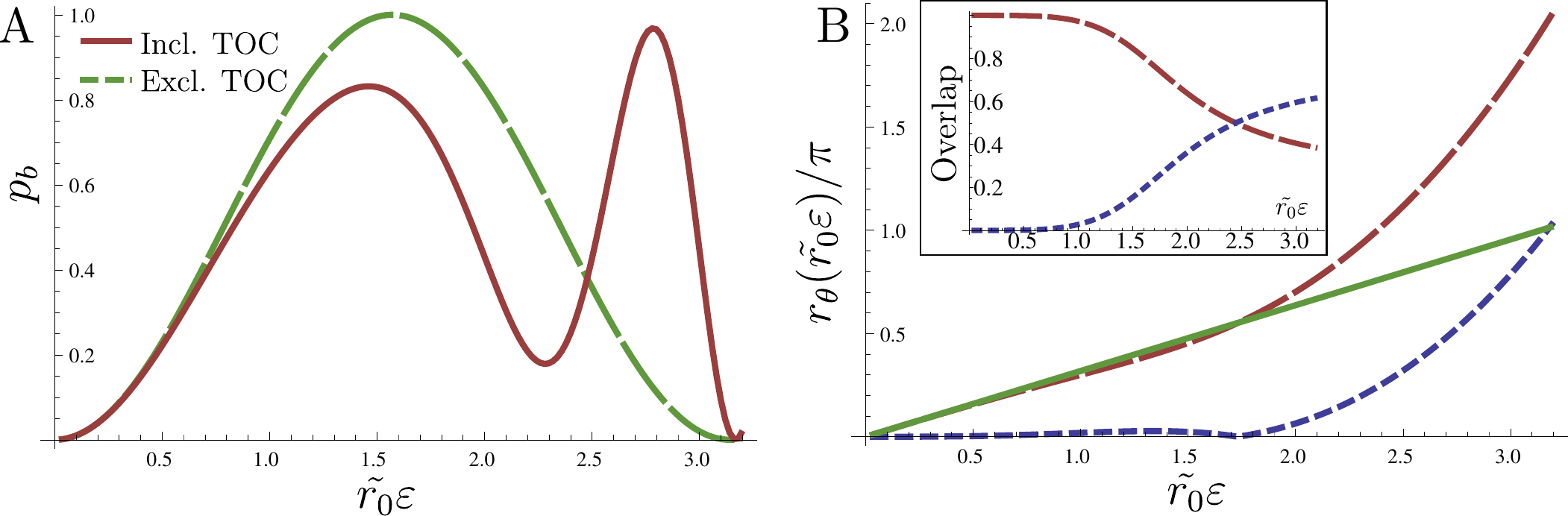}
\caption{\label{eff} In A we plot the probability of upconversion as a function of the interaction strength $\tilde r_0 \varepsilon$, for a single photon in the Gaussian wave packet $f(\alpha)$ including and excluding TOCs. In B we plot the Schmidt numbers of $\bar J$ including (dashed lines) and excluding (full line) TOCs. In the inset we plot the overlap $|c_\theta|^2$ between the time ordered corrected Schmidt functions and the Gaussian profile of the incoming single photon.} 
\end{figure}

Now let us consider what would happen if a single photon with fixed Gaussian profile $\mathcal{\hat A}_0^\dagger \ket{\vac}$ were sent to the FC device. We look at the probability of upconversion as a function of $\tilde r_0 \varepsilon$, and plot this probability, including and excluding TOCs, in Fig. \ref{eff}.A.
If TOCs could be eliminated (for instance following the approach of Reddy \emph{et al.} \cite{reddy14}), we would get $p_b=\sin^2(\tilde r_0 \varepsilon)$; instead, we get a more complicated curve. 
The upconversion curve attains a maximum near the ideal $\tilde r_0 \varepsilon \approx \pi/2$ but only reaches $\sim$ 80\%, a value consistent with experiments \cite{ben15}. 
The reason for this is easy to understand if we look at Fig. \ref{eff}.B, where we plot the Schmidt numbers and the overlap of the Schmidt functions with the single photon profile $f(\alpha)$. 
We see that up to $\tilde r_0 \varepsilon \leq 1.47 \sim \pi/2$, the Schmidt numbers are not significantly modified by the TOCs.
Yet the Schmidt functions are significantly modified at this point, and thus do not completely overlap with the shape of the incoming photon. This can be seen in Fig. \ref{eff}.B, where we plot the overlap of the Gaussian wave packet with the two Schmidt functions. 
As we increase $\tilde r_0 \varepsilon$ further, the probability of upconversion oscillates faster; this is just a reflection of the fact that past $\tilde r_0 \varepsilon=\pi/2$ the Schmidt numbers \emph{are} modified by the TOCs and grow much faster than linearly, as seen in Fig. \ref{eff}.B. This feature could be used as direct experimental evidence of the contribution of the TOCs to the JCA in FC.  
Our results suggest that in some FC experiments an increased pump power would eventually lead to enhanced upconversion rates, but only after an initial drop in the upconversion probability.
The final, most interesting feature of Fig. \ref{eff}.A is that the probability reaches unity for a value near $ \tilde r_0 \varepsilon \sim$  2.79. This happens because the two non-zero Schmidt numbers of $\bar J$ become odd multiples of $\pi/2$  at the same time. This allows the two Schmidt functions to cooperate and attain 100\% efficiency. The extra Schmidt mode is, as mentioned before, generated by the TOCs; thus one can think of this enhanced upconversion probability as an interference effect between the first and third order Magnus terms. Note that this upconversion process will be partially mode selective, since it only involves two Schmidt modes while leaving the other modes untouched.

In this letter we have studied the effects of the time ordering corrections (TOCs) on frequency conversion (FC) using the Magnus expansion. We have considered conversion using a $\chi_2$ material --- quasi-phase-matching can easily included --- and, with correspondences presented earlier \cite{nico14}, this can be generalized to the use of $\chi_3$ materials as well. We have shown how the TOCs modify the ideal operation of a quantum pulse gate and how they can be used to achieve near unity FC using the extra Schmidt functions that are generated by them. Finally, we provided a simple scaling argument that explains the results of Reddy \emph{et al.} \cite{reddy14}, in which TOCs are attenuated by using cascaded FC devices.

The authors acknowledge support of the National Science and Engineering Research Council of Canada.


\end{document}